
\font\ssbig=cmss10 scaled \magstep1  
\font\bfb=cmbx12                     
\font\tfont=cmbxti10

\baselineskip=12pt


\newfam\vecfam

\textfont\vecfam=\tfont \scriptfont\vecfam=\seveni
\scriptscriptfont\vecfam=\fivei

\def\msun{{\rm\,M_\odot}}

\def\pc{{\rm\,pc}}
\def\gyr{{\rm\,Gyr}}

\def\gtorder{\mathrel{\raise.3ex\hbox{$>$}\mkern-14mu
             \lower0.6ex\hbox{$\sim$}}}
\def\ltorder{\mathrel{\raise.3ex\hbox{$<$}\mkern-14mu
             \lower0.6ex\hbox{$\sim$}}}

\newcount\eqnumber
\eqnumber=1
\def\new{{\rm\the\eqnumber}\global\advance\eqnumber by 1}
\def\eqnam#1{\xdef#1{\the\eqnumber}}
%
\newcount\fignumber
\fignumber=1
\def\newfig#1{{\rm\the\fignumber}\xdef#1{\the\fignumber}\global\advance\fignumber by 1}
\def\refto#1{$^{#1}$}           
\def\ref#1{Ref.~#1}                     
\def\Ref#1{#1}                          
\gdef\refis#1{\item{#1.\ }}                     
\def\beginparmode{\endmode
  \begingroup \def\endmode{\par\endgroup}}
\let\endmode=\par
\def\body{\beginparmode}
\def\head#1{                    
  \goodbreak\vskip 0.5truein    
  {\centerline{\bf{#1}}\par}
   \nobreak\vskip 0.25truein\nobreak}
\def\references                 
  {\head{References}            
   \beginparmode
   \frenchspacing \parindent=0pt \leftskip=1truecm
   \parskip=8pt plus 3pt \everypar{\hangindent=\parindent}}
\def\endreferences{\body}

\catcode`@=11
\newcount\r@fcount \r@fcount=0
\newcount\r@fcurr
\immediate\newwrite\reffile
\newif\ifr@ffile\r@ffilefalse
\def\w@rnwrite#1{\ifr@ffile\immediate\write\reffile{#1}\fi\message{#1}}

\def\writer@f#1>>{}
\def\referencefile{
  \r@ffiletrue\immediate\openout\reffile=\jobname.ref%
  \def\writer@f##1>>{\ifr@ffile\immediate\write\reffile%
    {\noexpand\refis{##1} = \csname r@fnum##1\endcsname = %
     \expandafter\expandafter\expandafter\strip@t\expandafter%
     \meaning\csname r@ftext\csname r@fnum##1\endcsname\endcsname}\fi}%
  \def\strip@t##1>>{}}

\def\citeall#1{\xdef#1##1{#1{\noexpand\cite{##1}}}}
\def\cite#1{\each@rg\citer@nge{#1}}	

\def\each@rg#1#2{{\let\thecsname=#1\expandafter\first@rg#2,\end,}}
\def\first@rg#1,{\thecsname{#1}\apply@rg}	
\def\apply@rg#1,{\ifx\end#1\let\next=\relax
\else,\thecsname{#1}\let\next=\apply@rg\fi\next}

\def\citer@nge#1{\citedor@nge#1-\end-}	
\def\citer@ngeat#1\end-{#1}
\def\citedor@nge#1-#2-{\ifx\end#2\r@featspace#1 
  \else\citel@@p{#1}{#2}\citer@ngeat\fi}	
\def\citel@@p#1#2{\ifnum#1>#2{\errmessage{Reference range #1-#2\space is bad.}%
    \errhelp{If you cite a series of references by the notation M-N, then M and
    N must be integers, and N must be greater than or equal to M.}}\else%
 {\count0=#1\count1=#2\advance\count1 by1\relax\expandafter\r@fcite\the\count0,%
  \loop\advance\count0 by1\relax
    \ifnum\count0<\count1,\expandafter\r@fcite\the\count0,%
  \repeat}\fi}

\def\r@featspace#1#2 {\r@fcite#1#2,}	
\def\r@fcite#1,{\ifuncit@d{#1}
    \newr@f{#1}%
    \expandafter\gdef\csname r@ftext\number\r@fcount\endcsname%
                     {\message{Reference #1 to be supplied.}%
                      \writer@f#1>>#1 to be supplied.\par}%
 \fi%
 \csname r@fnum#1\endcsname}
\def\ifuncit@d#1{\expandafter\ifx\csname r@fnum#1\endcsname\relax}%
\def\newr@f#1{\global\advance\r@fcount by1%
    \expandafter\xdef\csname r@fnum#1\endcsname{\number\r@fcount}}

\let\r@fis=\refis			
\def\refis#1#2#3\par{\ifuncit@d{#1}
   \newr@f{#1}%
   \w@rnwrite{Reference #1=\number\r@fcount\space is not cited up to now.}\fi%
  \expandafter\gdef\csname r@ftext\csname r@fnum#1\endcsname\endcsname%
  {\writer@f#1>>#2#3\vskip -0.7\baselineskip\par}}

\def\ignoreuncited{
   \def\refis##1##2##3\par{\ifuncit@d{##1}%
     \else\expandafter\gdef\csname r@ftext\csname r@fnum##1\endcsname\endcsname%
     {\writer@f##1>>##2##3\vskip -0.7\baselineskip\par}\fi}}

\def\r@ferr{\endreferences\errmessage{I was expecting to see
\noexpand\endreferences before now;  I have inserted it here.}}
\let\r@ferences=\references
\def\references{\r@ferences\def\endmode{\r@ferr\par\endgroup}}

\let\endr@ferences=\endreferences
\def\endreferences{\r@fcurr=0
  {\loop\ifnum\r@fcurr<\r@fcount
    \advance\r@fcurr by 1\relax\expandafter\r@fis\expandafter{\number\r@fcurr}%
    \csname r@ftext\number\r@fcurr\endcsname%
  \repeat}\gdef\r@ferr{}\endr@ferences}


\let\r@fend=\endpaper\gdef\endpaper{\ifr@ffile
\immediate\write16{Cross References written on []\jobname.REF.}\fi\r@fend}

\catcode`@=12

\citeall\refto		
\citeall\ref		%
\citeall\Ref		%

\centerline{\bfb Searching for the Microlenses: The Observational Signatures of Old White Dwarfs}

\medskip
\centerline{\bfb Brad~M.S.~Hansen }
\smallskip    
Canadian Institute for Theoretical Astrophysics, University of Toronto,
Toronto, ON, M5S 3H8, Canada
\bigskip

{\ssbig
The recent discovery of microlensing of stars in the Large Magellanic
Cloud\refto{Macho1,Eros1} has excited much interest in the nature of the
lensing population. Detailed analyses indicate that the mass of these
objects ranges from $0.3-0.8 \msun$\refto{Macho2}, suggesting that
they might be white dwarfs, the faint remnants of stellar evolution. The confirmation
of such an hypothesis would yield profound insights into the early history of
our galaxy and the early generations of stars in the universe\refto{Larson,Charlot,Adams}.
Previous attempts have been made to place theoretical constraints on this
scenario\refto{Adams,Chab,Graff}, but were unduly pessimistic because they
relied on inadequate evolutionary models.
Here we present the first
results from detailed evolutionary models appropriate for the study of white dwarfs of truly
cosmological vintage.
 We find that the commonly held notion that old white dwarfs are red
to hold only for helium atmosphere dwarfs and that hydrogen atmosphere dwarfs will
be blue, with colours similar to those of the faint point sources found in the Hubble
Deep Field.
 Thus, any direct
observational search for
the microlensing population should search for faint blue objects rather than faint red
ones.

}
\medskip

The issue of old white dwarf observability is inextricably linked to the age of the
object, since white dwarfs fade with time. The
primary uncertainty in the calculation of accurate cooling models is 
the description of energy transport in the stellar
atmosphere, which affects both the cooling rate\refto{DM} and observational appearance\refto{BSW}.
Old white dwarfs have luminosities $\log L/L_{\odot} < -4$ and effective temperatures $T_{\rm eff} < 6000$~K.
The atmospheric constituents at these temperatures are neutral, so that the primary opacity
sources are the collisionally induced absorption of molecular hydrogen (for hydrogen atmospheres) or
Rayleigh scattering (for helium atmospheres). The molecular opacities are strongly wavelength
dependent and require a detailed radiative transfer calculation. The calculation of such
atmosphere models\refto{BSW} has led to advances in the study of the basic physical parameters of old
white dwarfs\refto{BRL}. However, previous cooling models have used outer boundary conditions calculated
using simplified atmosphere calculations, so that the self-consistent determination of cooling ages
from atmospheric models was not possible for the oldest white dwarfs. The cooling models described here aim
to rectify that uncertainty.

Our calculations use a white dwarf cooling code originally developed for the study of the companions to
millisecond pulsars\refto{HP1}. To this we have added a
 detailed atmospheric model using the Feautrier and Avrett-Krook
methods\refto{Mihalas} to solve the radiative transfer at the surface. The
 atmospheric model provides an outer boundary condition (taken to be the photosphere) for the
 interior model which describes the cooling of the white dwarf.
 The position of the photosphere for helium and hydrogen atmospheres differs dramatically because
 neutral helium does not form molecules
 so that the opacity $\kappa$ is much lower
and hence the density at the photosphere $\rho \propto 1/\kappa T_{\rm eff}$ is much larger.
Figure~1 shows the location
 of the photosphere for
hydrogen and helium atmospheres respectively.
In both cases the photospheric density increases as
the star cools and $T_{\rm eff}$ drops. This trend ends when both atmospheres reach $T_{\rm eff} \sim$~3000~K,
but for different reasons. For helium atmospheres, pressure ionization of helium increases the opacity
 dramatically,
halting the inward motion of the photosphere. For hydrogen atmospheres, the molecular hydrogen opacity
is stronger at long wavelengths $\lambda > 1 \rm \mu m$, so that the opacity increases when the peak of the
black-body $\lambda_{\rm bb} \sim 1.7 {\rm \mu m}/ (T_{\rm eff}/3000 {\rm K})$ moves into that range, and hence the
photospheric density moves outwards again. The molecular opacity is strongly wavelength-dependent, so
that the observational appearance of these old stars deviates significantly from a black-body spectrum,
in a similar fashion to that seen in brown dwarf atmospheres\refto{Saumon}.

Figure~2 shows the comparison of our cooling results with the  best extant models in the 
literature\refto{Wood95,Salaris}.
 The agreement with the hydrogen atmosphere models
is excellent until the models reach $T_{\rm eff} \sim 6000$~K, at which point our improved treatment of
the atmospheric physics results in slower cooling. The difference in the helium models is more dramatic,
with the new models cooling rather more rapidly. This is a result of the extremely low opacity of
neutral helium.
Application of the new models to the white dwarf luminosity function
 shows that previous age estimates for disk hydrogen dwarfs 
were reasonably accurate, although disk helium dwarf ages were somewhat overestimated. However,
for white dwarfs residing in the galactic halo or in globular clusters, the differences
are very important. In particular, previous efforts to place constraints on white dwarf
dark matter\refto{Adams,Chab,Graff} have been unduly pessimistic because they use inappropriate
white dwarf models.

The observational constraints on local white dwarf dark matter centre on two sources, the luminosity
function determined from proper motions\refto{LDM} and searches for point sources in
the Hubble Deep Field\refto{FGB,ESG,MM}. Unfortunately, the proper motion sample
suffers from poorly constrained incompleteness at the fainter magnitudes
(the luminosity function declines much more rapidly than the luminosity function based on
white dwarfs in binaries\refto{Oswalt}). Thus, inferring how many objects could have been
seen at fainter magnitudes is a particularly dangerous endeavour. More robust estimates must rely
on the Hubble Deep Field (HDF) alone.
Figure~3 shows the point sources detected by various groups in the
HDF along with cooling tracks for both hydrogen and helium atmospheres.
The first result that one can glean from this is that the HDF places no limits on helium atmosphere
dwarfs, because the rapid cooling means that they become unobservable within $6 \gyr$, which is
approximately half their expected age. The second and most interesting result is that the hydrogen
atmosphere dwarfs are potentially observable and would lie in the region of the faint blue objects
that were indeed detected. This is contrary to the conventional wisdom which
states that white dwarfs should become redder with age because that is the trend shown by a
black body (and indeed that is what happens to the helium dwarfs). The blueward deviation shown
by the hydrogen atmosphere dwarfs is again a consequence of the strong molecular hydrogen opacity
which also causes the photosphere to move to lower densities in Figure~1.
The strong
opacity at longer wavelengths (red) forces the stellar flux to emerge at shorter wavelengths (blue)
where the opacity is smaller. Thus, hydrogen atmosphere white dwarfs may have already been detected
in the halo!

To quantitatively connect these findings to the observational searches one also requires a model for the
relative populations of hydrogen and helium atmosphere white dwarfs. Recent advances in 
asteroseismological investigations of white dwarfs\refto{Clemens} have confirmed that many hydrogen
atmosphere white dwarfs have hydrogen surface layers of mass $\sim 10^{-4} \msun$. This
suggests that such stars will survive as hydrogen atmosphere dwarfs despite the mixing effects
of surface convection zones. Although these estimates are for disk white dwarfs, the mechanisms
for removing hydrogen from the surface of a white dwarf\refto{DM,IM,Pac} should become less
efficient in stars with lower metallicity, such as those in the halo and in globular clusters.
The relative proportion of hydrogen atmospheres amongst the cool white dwarfs in the disk
appears to be $>50\%$, so we will adopt this
as a conservative estimate. In the future, the effect of advanced age and lower progenitor metallicity can
be empirically tested by examining the faint white dwarfs in globular clusters\refto{Richer}.

How many white dwarfs would we expect to see in the Hubble Deep Field?
The HDF probes great distances, but only over a very small portion of sky, so that
the total galactic volume sampled is quite small. Given a limiting magnitude $m_V \sim 28$ and
an absolute magnitude for old white dwarfs $M_V \sim 17$, the corresponding volume probed is
only
\eqnam{\VHDF}$$ V_{\rm HDF} \sim (5.4 \pc)^3 10^{0.6 ((m_V-28) - (M_V-17))}.
\eqno(\new)
$$
Thus, even if the white dwarfs constitute the entire local dark matter
($\sim 0.01 \msun \pc^{-3}$)\refto{BSS}, their space density is only $\sim 0.02\pc^{-3}$ and hence 
we expect to find only 3 even if all the white dwarfs have hydrogen atmospheres and are 12 Gyr old
(an age appropriate for globular cluster and halo stars).
Our most realistic estimate invokes only $50 \%$
of the total number in hydrogen atmosphere white dwarfs and requires that only $50\%$ of the dark matter is in the form of white dwarfs\refto{Macho2}.
In this case,  
the volume probed by the HDF is too small to reliably detect even one object. Nevertheless, as one
can see in Figure~3, several points sources have been detected in the HDF.
 Given the relative profusion of these objects,
 most must have another origin, such as extragalactic star forming regions\refto{ESG}.

The implications of these results for the search for dark matter are twofold. The first result is
that deep surveys for very red objects are unlikely to be successful in the near future because the
 red population (helium atmosphere dwarfs) will be very faint. However, searches for faint blue objects could be much more
successful. The second implication is that the hydrogen atmosphere dwarfs are observable but that
the HDF results suggest that there are other unresolved populations of objects with similar colours and 
magnitudes. Thus, proper motions will be essential for identifying galactic objects from extragalactic
ones. 
\references
{\parskip=0pt
\refis{DM} D'Antona, F. \& Mazzitelli, I., Cooling of White Dwarfs, {\it Ann.\ Rev.\ Astr.\ Astrophys.},
 {\bf 28}, 
139-181, (1990) \par
\refis{BSW} Bergeron, P., Saumon, D. \& Wesemael, F., New model atmospheres for very cool white
dwarfs with mixed H/He and pure He compositions, {\it Astrophys.\ J.}, {\bf 443}, 764-779 (1995) \par
\refis{Mihalas} Mihalas, D., {\it Stellar Atmospheres}, W.H.Freeman \& Co., San Francisco, (1970) \par
\refis{Saumon} Saumon, D., Bergeron, P., Lunine, J. I., Hubbard, W. B. \& Burrows, A., Cool zero-metallicity
stellar atmospheres, {\it Astrophys.\ J.}, {\bf 424}, 333-344 (1994) \par
\refis{Wood95} Wood, M. A., Theoretical White Dwarf Luminosity Functions: DA Models', in {\it White Dwarfs},
(Koester, D. \& Werner, K., eds), 41 (Springer-Verlag) (1995) \par
\refis{Salaris} Salaris, M., {\it et al},
The cooling of CO white dwarfs: Influence of the internal chemical distribution', {\it Astrophys.\ J.}, {\bf 486}, 413-419 (1997) \par
\refis{LDM} Liebert, J., Dahn, C. D. \& Monet, D. G., The luminosity function of white dwarfs, {\it Astrophys.\ J.}, {\bf 332}, 891-909 (1988) \par
\refis{Oswalt} Oswalt, T. D., Smith, J. A., Wood, M. A. \& Hintzen, P., A lower limit of 9.5 gyr on the
age of the galactic disk from the oldest white dwarf stars, {\it Nature}, {\bf 382}, 692-694 (1996) \par
\refis{Clemens} Clemens, J. C., The pulsation properties of the DA white dwarf variables, {\it Baltic Ast.}, {\bf 2}, 407-434 (1993) \par
\refis{IM} Iben, I.A \& MacDonald, J., The Effects of Diffusion due to gravity and due to composition
gradients on the rate of hydrogen burning in a cooling degenerate dwarf. I. The case of a thick helium buffer
layer, {\it Astrophys.\ J.}, {\bf 296}, 540-553 (1985) \par
\refis{Pac} Paczynski, B., Evolution of Single Stars. VI. Model Nuclei of Planetary Nebulae, {\it Acta Astron.}, {\bf 21}, 417-435 (1971) \par
\refis{BSS} Bahcall, J. N., Schmidt, M. \& Soneira, R. M., The Galactic Spheroid, {\it Astrophys.\ J.}, {\bf 265}, 730-747
 (1983) \par
\refis{Macho1}Alcock, C. {\it et al}, Possible gravitational microlensing of a star in the Large
Magellanic Cloud,  {\it Nature}, {\bf 365}, 621-623 (1993)\par
\refis{Eros1} Aubourg, E. {\it et al}, Evidence for gravitational microlensing by dark objects in the galactic
halo, {\it Nature}, {\bf 365}, 623-625 (1993)\par
\refis{Macho2}Alcock, C. {\it et al}, The MACHO project Large Magellanic Cloud Microlensing Results
from the First two years and the nature of the galactic dark halo, {\it Astrophys.\ J.} {\bf 486}, 697-726 (1997)  \par
\refis{Richer} Richer, H. B., {\it et al}, White dwarfs in globular clusters: Hubble Space Telescope observations of M4, {\it Astrophys.\ J.} {\bf 484}, 741 (1997) \par
\refis{Charlot} Charlot, S. \& Silk, J., Signatures of White dwarf galaxy halos, {\it Astrophys.\ J.},
{\bf 445}, 124-132 (1995) \par
\refis{Adams} Adams, F. \& Laughlin, G., Implications of White dwarf galactic halos,
 {\it Astrophys.\ J.}, {\bf 468}, 586-597 (1996) \par
\refis{Chab} Chabrier, G., Segretain, L. \& Mera, D.,
Contribution of brown dwarfs and white dwarfs to recent microlensing observations and the halo
mass budget,
{\it Astrophys.\ J.}, {\bf 468}, L21-L24 (1996) \par
\refis{HP1} Hansen, B. M. S. \& Phinney, E. S., Stellar Forensics I: Cooling curves, 
{\it Mon.\ Not.\ R.\ Astron.\ Soc.}, {\bf 294}, 557-568 (1998) \par
\refis{Graff} Graff, D. S., Laughlin, G. \& Freese, K.,
MACHOs, white dwarfs and the age of the universe,
{\it Astrophys.\ J.}, {\bf 499}, 7-19 (1998) \par
\refis{Larson} Larson, R. B.,  Dark matter - Dead stars?, {\it Com.\ Astrophys.}, {\bf 11}, 273-282, 1987
\par
\refis{FGB} Flynn, C., Gould, A. \& Bahcall, J. N.,
Hubble Deep Field constraint on baryonic dark matter,
 {\it Astrophys.\ J.}, {\bf 466}, L55-L58 (1996) \par
\refis{ESG} Elson, R. A. W., Santiago, B. X., \& Gilmore, G. F.,
Halo stars, starbursts, and distant globular clusters: A survey of unresolved objects in the Hubble
Deep Field,
{\it New Astron.}, {\bf 1}, 1-16 (1996) \par
\refis{MM} Mendez, R. A., Minniti, D., de Marchi, G., Baker, A. \& Couch W. J.,
Starcounts in the Hubble Deep Field: Constraining Galactic Structure Models.,
 {\it Mon.\ Not.\ R.\ Astron.\ Soc.}, {\bf 283},
666-672 (1996) \par
\refis{BRL}
 Bergeron, P., Ruiz, M. T., \& Leggett, S. K., The chemical evolution of cool white dwarfs
and the age of the local galactic disk, {\it Astrophys.\ J.\ Suppl.}, {\bf 108}, 339-387, (1997) \par
\refis{Chaboyer} Chaboyer, B., DeMarque, P., Kernan, P. J. \& Krauss, L. M., The Age of Globular Clusters
in the light of Hipparcos: Resolving the Age Problem?, {\it Astrophys.\ J.}, {\bf 494}, 96-110 (1998) \par
\refis{Holtz} Holtzman, J. A. {\it et al}, The Photometric Performance \& Calibration of WFPC2, {\it Publ.\ Ast
ron.\
Soc.\ Pac.}, {\bf 107}, 1065-1093 (1995) \par
}
\endreferences
\endmode
\bigskip
\par\noindent ACKNOWLEDGEMENTS. The author thanks the Aspen Center for Physics for hospitality while
some of the ideas in this paper were first germinated. Comments by Drs Wood, Chabrier, Garcia-Berro \&
Hernanz are gratefully acknowledged.

\medskip
Correspondence to B. Hansen (email: hansen@cita.utoronto.ca)

\bigskip

Caption: Figure~1

{\bf Location of the Photosphere:} The solid and dotted lines
indicate the evolution of the photospheres for hydrogen and helium
atmospheres respectively. The solid and open circles are the same
quantities from the work of Bergeron et al\refto{BSW}. 

\medskip

Caption: Figure~2

{\bf Cooling curves:} 
a) The solid line from this work is compared with the dashed line from the models of Wood (1995)
\refto{Wood95}.
The model is for a 0.6$\msun$ white dwarf with an oxygen core and a hydrogen mass fraction of $10^{-4}$.
The improved boundary condition leads to a slight flattening of the cooling curve at $\log L/L_{\odot}
\sim -4$ and longer cooling times thereafter.
b) The solid line from this work is compared with the dashed line from the models of 
Salaris et al (1997)\refto{Salaris}.
The model is for a $0.6 \msun$ carbon/oxygen model with a helium mass fraction of $10^{-3.25}$.
The much longer cooling of the Salaris model is because of the outdated atmospheric 
model
used in that work. In both panels the horizontal dotted lines enclose the location of the turnover in
the disk white dwarf luminosity function. The vertical dotted lines indicate the mean age of the
globular clusters\refto{Chaboyer}, indicating the expected age of any halo white dwarf population.
The large variations in the different models at these ages indicate how important accurate 
atmosphere models are.

\medskip

Caption: Figure~3

{\bf Stellar Objects in the Hubble Deep Field:} The filled circles are
the unresolved objects detected by Elson et al\refto{ESG}. The open circles are the point sources from
Mendez et al\refto{MM}. Although Flynn et al\refto{FGB} did not publish a table of detections, the dotted line
encloses their `halo region', which they used to constrain the halo white dwarf population.
 The dashed lines show the cooling behaviour of a helium atmosphere white
dwarf at a distance of 1 kpc (upper line) and 2 kpc (lower line). The upper curve is labelled
with ages in Gyrs at appropriate points. The solid lines represent the corresponding evolution of
a  hydrogen atmosphere dwarf at 1 and 2 kpc. The dwarf ages are shown on the lower curve in this
case. The bandpasses are the hubble telescope bandpasses from
 Holtzmann et al\refto{Holtz}. The blueward shift for ages 10-12 Gyr is
due to the presence of molecular hydrogen in the atmosphere.
\vfill\eject

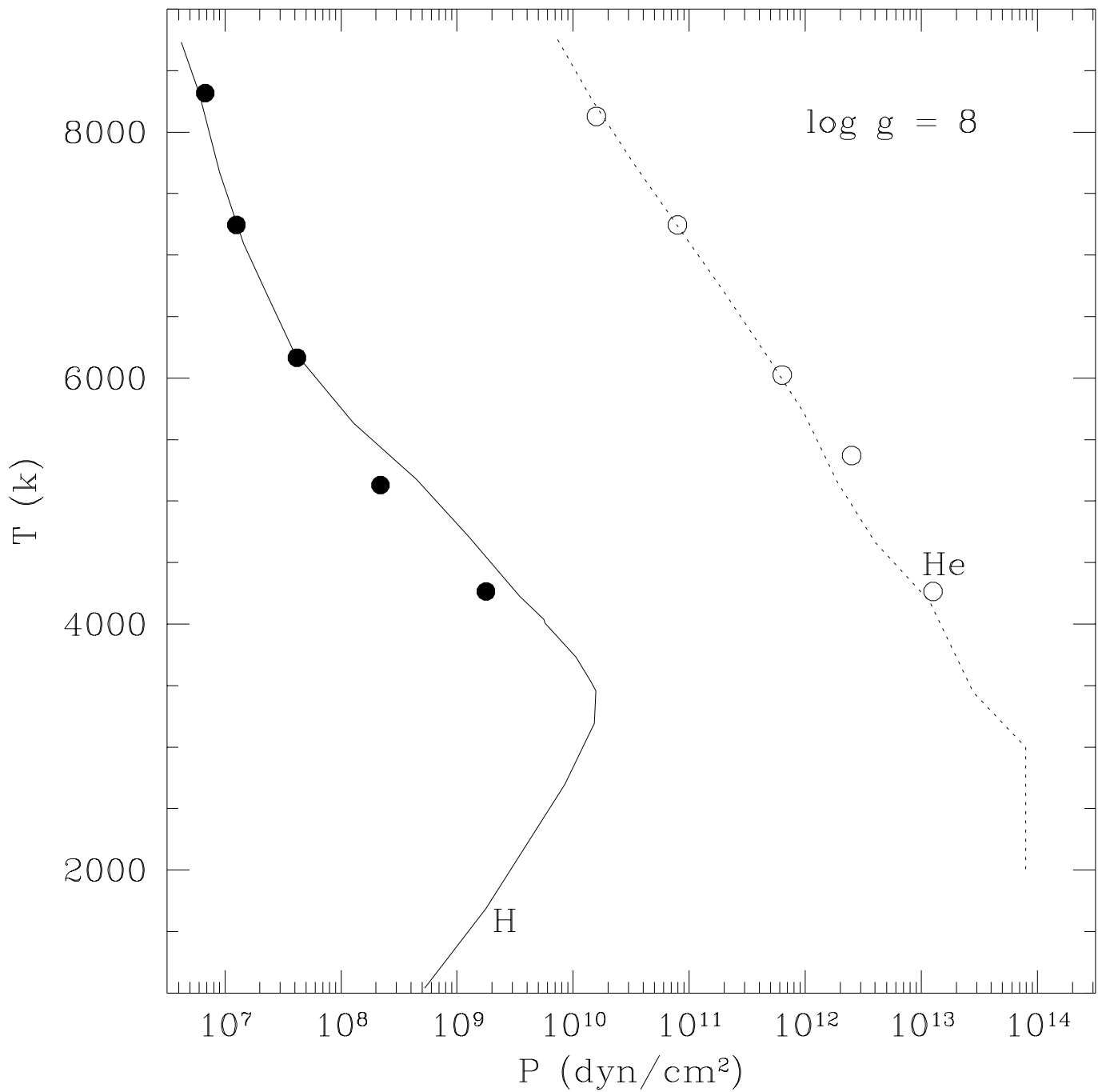
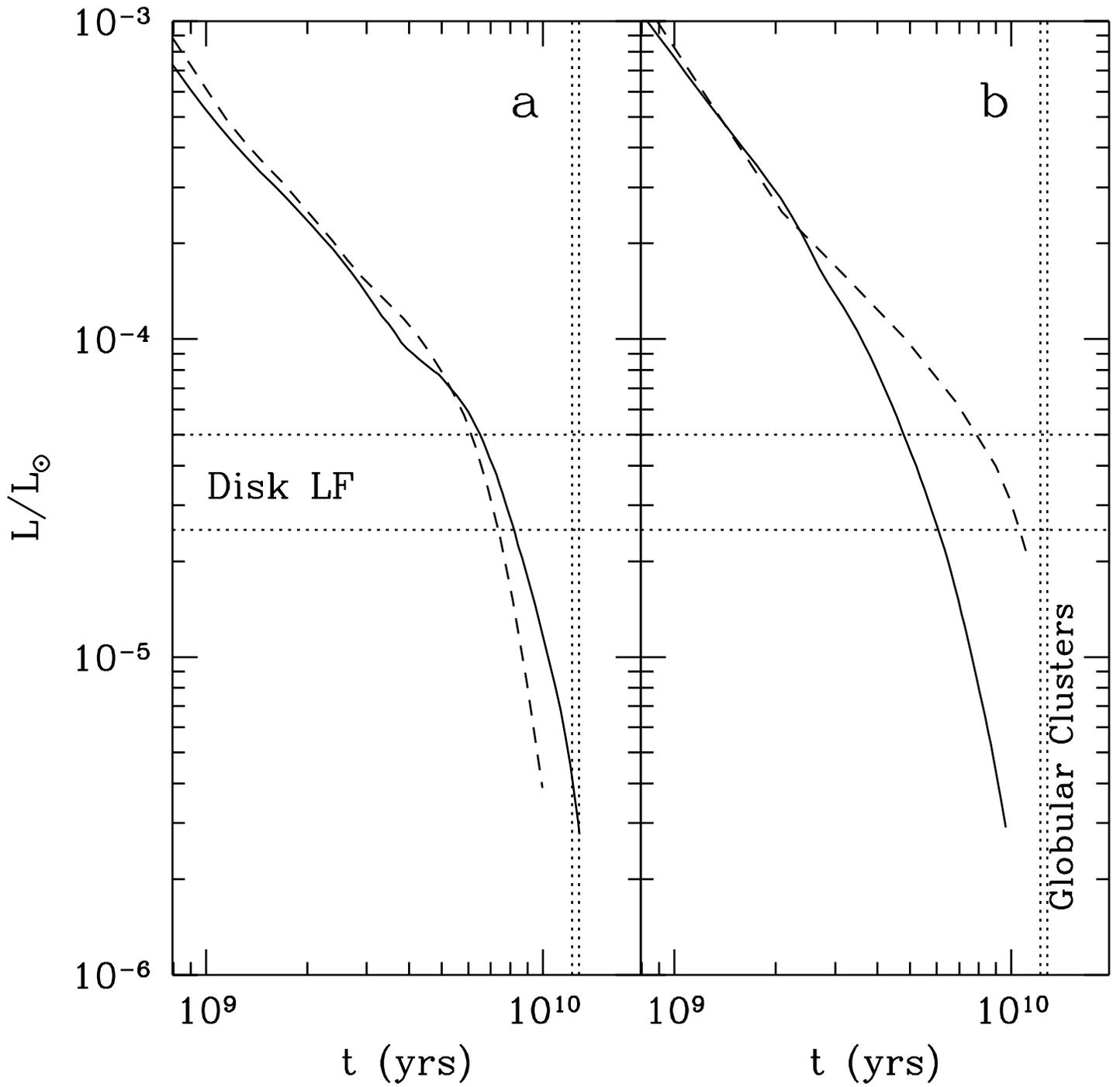
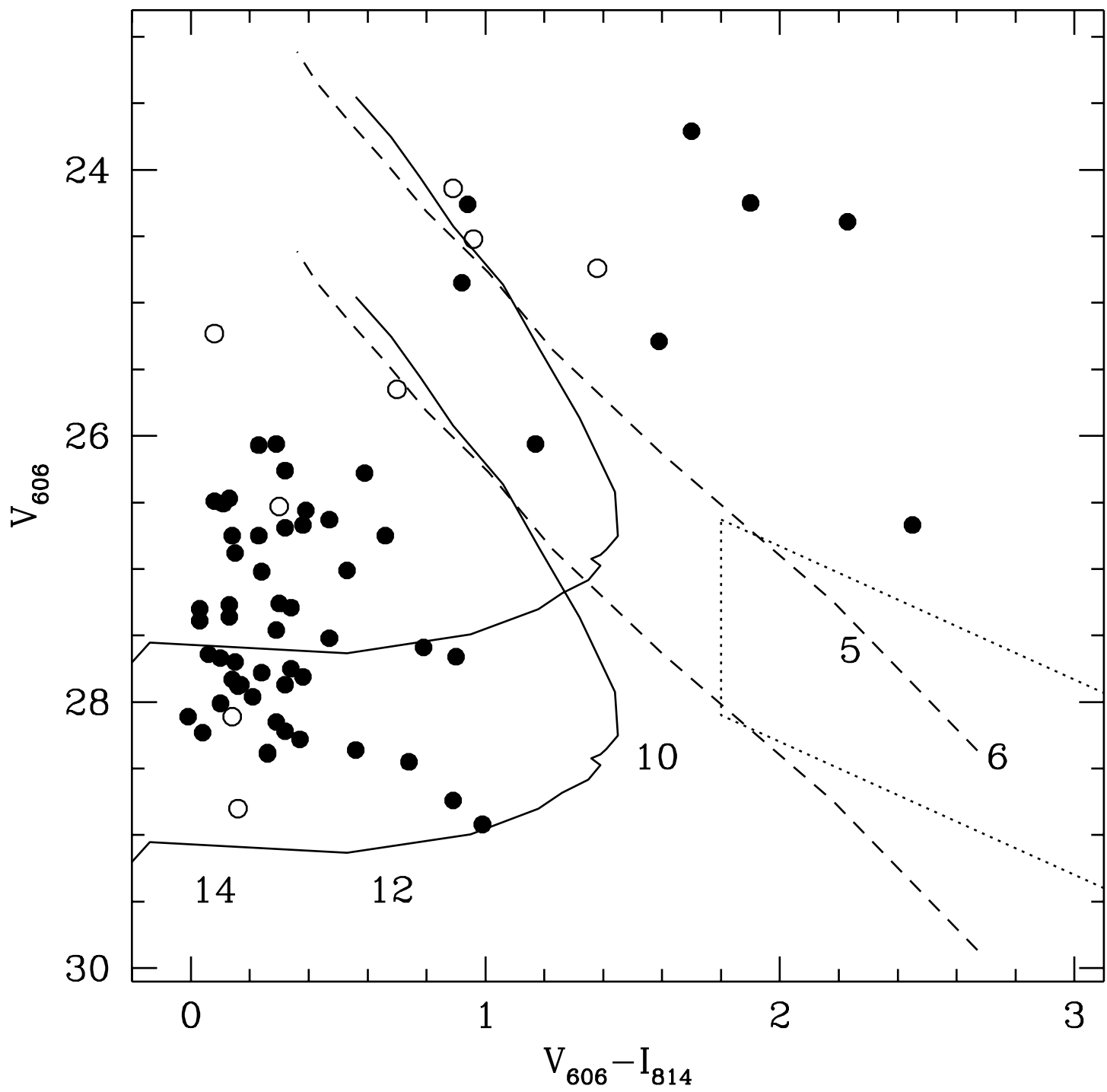
\bye